\documentclass[aps,pra,longbibliography,twocolumn,superscriptaddress]{revtex4-1} 

\usepackage{graphicx}
\usepackage[hidelinks]{hyperref}

\usepackage[percent]{overpic} 
\usepackage{xcolor} 
\usepackage{upgreek}
\usepackage{ulem}

\newcommand{\textold}[1]{\textcolor{red!90!black}{\sout{#1}}}
\newcommand{\textnew}[1]{\textcolor{green!50!black}{#1}}

\renewcommand{\textold}[1]{}
\renewcommand{\textnew}[1]{#1}

\begin{document}
    
    \title{Transient supersolid properties in an array of dipolar quantum droplets}
    
    
    \author{Fabian B\"{o}ttcher}
    \author{Jan-Niklas Schmidt}
    \author{Matthias Wenzel}
    \author{Jens Hertkorn}
    \author{Mingyang Guo}
    \author{Tim Langen}
    \author{Tilman Pfau}
    \email{t.pfau@physik.uni-stuttgart.de}
    \affiliation{5{.} Physikalisches Institut and Center for Integrated Quantum Science and Technology, Universit{\"a}t Stuttgart, Pfaffenwaldring 57, 70569 Stuttgart, Germany}

    \date{\today}

\begin{abstract}
We study theoretically and experimentally the emergence of supersolid properties in a dipolar Bose-Einstein condensate. The theory reveals a ground state phase diagram with three distinct regimes -- a regular Bose-Einstein condensate, incoherent and coherent arrays of quantum droplets. The coherent droplets are connected by a \textold{finite superfluid density background}\textnew{background condensate}, which leads -- in addition to the periodic density modulation - to a robust phase coherence throughout the whole system. We further theoretically demonstrate that we are able to dynamically approach the ground state in our experiment and that its lifetime is only limited by three-body losses. Experimentally we probe and confirm the signatures of the phase diagram by observing the in-situ density modulation as well as the phase coherence using matter wave interference.
\end{abstract}
    
\maketitle

Whether a material is solid, liquid or gaseous in classical physics depends on the strength of the interactions with respect to the motional energy of the particles. In analogy to this behavior the interplay between quantum fluctuations and interparticle interactions also leads to new phases of matter in quantum mechanics. One example of such quantum phases is the supersolid \cite{Boninsegni2012, Leggett1970, Andreev1969, Thouless1969, Chester1970, Penrose1956} featuring the periodic density modulation of a solid together with the dissipationless flow of a superfluid. While these properties are normally thought of as mutually exclusive, it was shown that they can actually coexist \cite{Leggett1970}. More formally speaking a supersolid features both on- and off-diagonal long-range order in its density matrix \cite{Boninsegni2012}.  Originally, supersolidity was mainly discussed in the context of $^{4}$He, for which it remains elusive in the experiments \cite{Chan2013}. The concept of supersolidity has since been generalized to other superfluid systems and supersolid properties have been observed in ultracold atomic systems for spin-orbit coupled Bose-Einstein condensates (BECs) \cite{Li2017} as well as BECs symmetrically coupled to two crossed optical cavities \cite{Leonard2017, Leonard2017a}. In these systems the periodicity of the modulation is induced by the underlying periodic optical potentials. In contrast, there are physical systems where the self-organized structure formation is induced by the intrinsic interactions and therefore phonon modes of the periodic modulation are allowed like in classical solids. 

One promising system of \textold{the} this type are dipolar quantum gases \textnew{\cite{Baranov2012, Lu2015}}, featuring both short-range contact interactions as well as long-range dipole-dipole interactions. These dipolar systems feature a rotonic dispersion relation similar to $^{4}$He \cite{Santos2003} which in addition, is fully tunable by changing the contact interaction strength as well as the external confinement along the dipoles. This dispersion relation has been studied experimentally \cite{Chomaz2018, Petter2018} and has led to the discovery of 2D arrays of quantum droplets \textnew{\cite{Kadau2016, Ferrier-Barbut2016, Petrov2015}}. However, it was shown that the 2D arrays in these early experiments were incoherent, excited states of the system. For the considered geometries the ground state was always a single droplet \cite{Wachtler2016, Bisset2016,Chomaz2016, Ferrier-Barbut2018}. In contrast to this, in \cite{Wenzel2017} we pointed out that the ground state in strongly-confined 2D geometries is made up of droplet arrays, but experimentally observed that these arrays rapidly loose their \textnew{relative} coherence during their dynamical formation process. \textnew{So while each droplet is coherent by itself, there is no global phase coherence between different droplets. In this work we always refer to this global coherence of the system.} Furthermore, increasing the overlap of the droplet wavefunctions through an increase of the remaining weakly confining trapping direction was proposed as a way to establish more robust phase coherence in experiments.

A recent theoretical study \cite{Roccuzzo2018} examined a similar elongated trapping configuration with periodic boundaries along one axis perpendicular to the magnetic field. In this work it was shown that close to the softening of the roton mode droplets form, which are immersed in the dilute superfluid background of a BEC. This special case of coherent droplets forms only in a very narrow range of interaction strengths, while for smaller contact interaction strengths the background vanishes, leading to isolated droplets. The coherent droplets were shown to exhibit two distinct excitation modes, a phonon and a phase mode, which are hallmarks of supersolidity.

\textold{Shortly afterwards it was indeed} \textnew{At the same time it was} observed experimentally that phase coherent droplets can exist for a narrow range of contact interaction strengths \cite{Tanzi2018}. However, in this experiment only the phase information after time-of-flight expansion was accessible \footnote{During the writing of this work we became aware of another related time-of-flight investigation with Er atoms. Private Communication, F. Ferlaino} and a detailed theoretical explanation of the observations is still lacking.

\begin{figure}[t]
\begin{overpic}[width=0.48\textwidth]{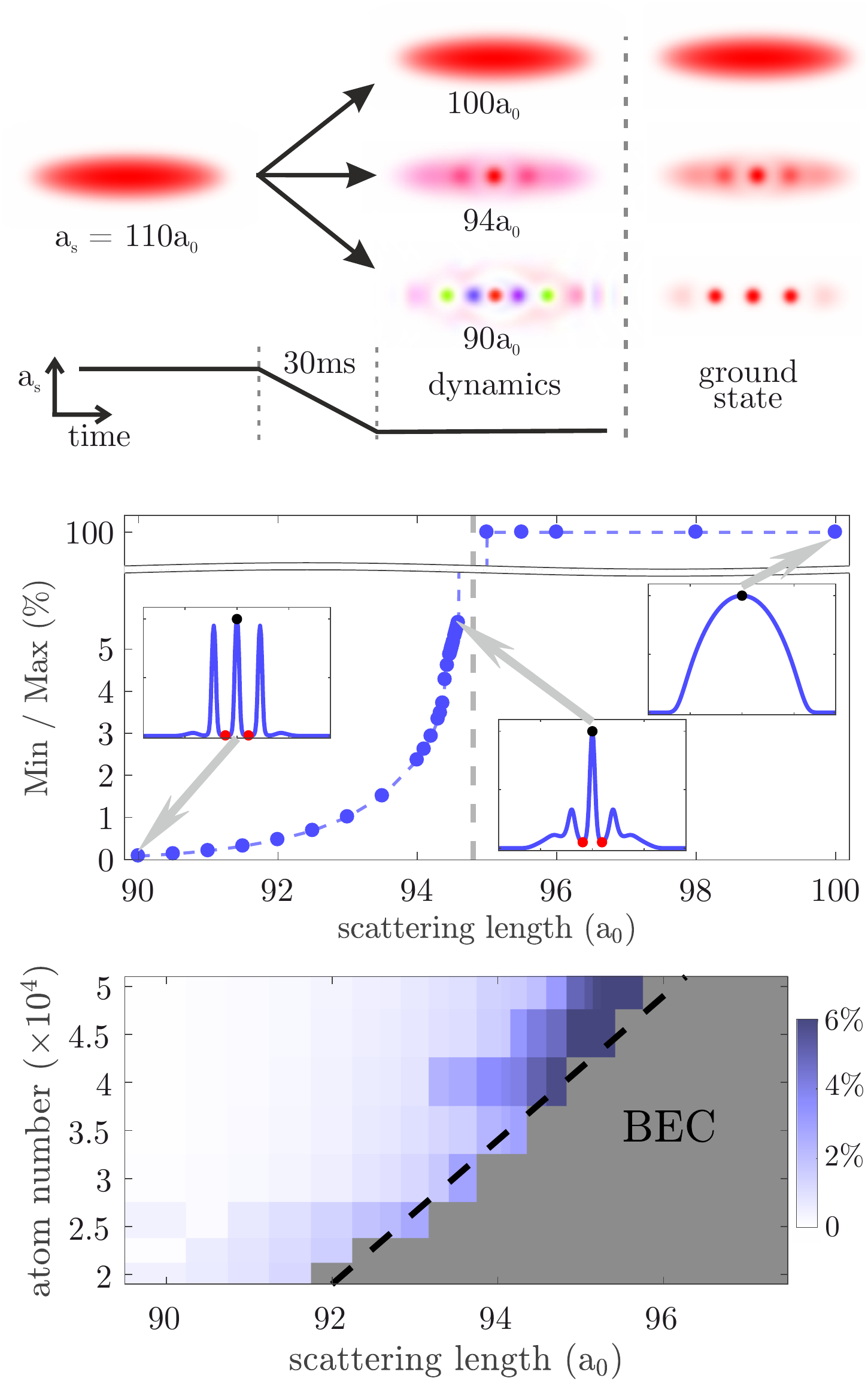}
\put(0,94){a)} \put(0,62){b)} \put(0,31){c)}
\end{overpic}
\caption{\label{fig:1} 
\textbf{Phase transition from BEC to immersed droplets to isolated droplets.} a) Schematic of the experimental sequence and corresponding ground states. We start from a BEC and then change the scattering length to its final value, where we observe 3 different regimes: at high scattering lengths a BEC, at low scattering lengths isolated droplets and in between droplets immersed in a superfluid background. We compare the dynamical simulations to the corresponding ground states. The pictures show the phase in color scale weighted by the density distribution. b) To quantify the transition we calculate the ratio of the first minimum compared to the center peak height of the ground state containing $3.5 \times 10^4$ atoms. c) Ground state phase diagram of this calculated ratio for different atom numbers and scattering lengths. Note that with higher atom number the number of droplets, and, as a consequence, also their overlap increase. The dashed black line is a guide to the eye.
}
\end{figure}

Here we present a comprehensive study of the supersolid properties of a trapped dipolar gas. First, we show theoretically that within the framework of the extended Gross-Pitaevskii equation (eGPE) a phase-coherent and density-modulated state can be reached dynamically for our elongated trap geometry. This state is found to be very close to the ground state of the system. Second, we experimentally realize such a state and observe its properties, both in-situ and in time-of-flight. With these two complementary observation techniques we map out the signatures of the theoretical phase diagram that clearly reveals both a coherent and an incoherent density-modulated regime. 

We start by looking at the theoretical ground state of the dipolar system within the framework of the eGPE. To this end, we numerically solve the eGPE using imaginary time propagation \cite{Wenzel2017} to calculate the ground states. These simulations are performed for $2-5 \times 10^4$ $^{162}$Dy atoms in a harmonic potential with trapping frequencies $\omega~=~2\pi\,$(18.5,~53,~81)\,Hz similar to \cite{Tanzi2018}. Depending on the contact interaction strength this yields three distinct regimes, which are summarized on the right side of Fig. \ref{fig:1}a for an atom number of $3.5\,\times\,10^4$. While for high scattering lengths of $a_{\text{s}} \geq 95\,a_{\text{0}}$ the ground state is a regular BEC, at low scattering lengths $a_{\text{s}} \lesssim 90\,a_{\text{0}}$ we recover the thoroughly studied regime of isolated droplets \cite{Kadau2016, Ferrier-Barbut2016,Ferrier-Barbut2016a, Wenzel2017}. The most interesting regime, e.g. for $a_{\text{s}} = 94\,a_{\text{0}}$, is found in between these two limits and consists of droplets that are immersed in a residual BEC background. This background acts as a link between the droplets and therefore establishes phase coherence.

With these observations in mind we quantify the transition in terms of the density link between the individual droplets, in particular between the central droplet and its nearest neighbors. To do this we analyze cuts through the center of the simulated densities $n = |\psi|^2$ and calculate the ratio between the first minimum and the central maximum. This simple measure characterizes the density overlap between neighboring droplets and is shown in Fig.~\ref{fig:1}b. For a BEC there exists no density modulation and therefore we set this ratio to 100\%. As soon as the density modulation emerges below $a_{\text{s}} \approx 94.5\,a_{\text{0}}$ this ratio is well-defined and shows a distinct jump followed by a steady decrease for lower scattering lengths. Interpreting the overlap as an order parameter this discontinuity is an indication of a first-order phase transition\textnew{, further evidenced by the observation of hysteresis in our numerical simulations \cite{SupMat} and by the experimental signatures in \cite{Tanzi2018}}. Moreover, this behavior is reminiscent of the decrease of the superfluid fraction across the supersolid phase transition that was observed in previous works \cite{Roccuzzo2018, Macri2013}. However, note that while the ratio shown in Fig.~\ref{fig:1}b is a measure of the overlap of the droplets, it does not directly correspond to the actual superfluid fraction. The superfluid fraction of the system was shown to notably exceed the values of this overlap ratio \cite{Roccuzzo2018}.

We extend this study to different atom numbers and show it as a phase diagram in Fig.~\ref{fig:1}c. We observe a clear phase boundary where the density modulated state becomes lower in energy than a regular BEC. This phase boundary shifts to higher scattering length with increasing atom number. While overall being located close to the roton instability our observed scaling for the simulated ground states appears different from the approximate roton scaling for a trapped gas reported in \cite{Tanzi2018}, in particular at higher interaction strengths.

\begin{figure}[t]
\begin{overpic}[width=0.48\textwidth]{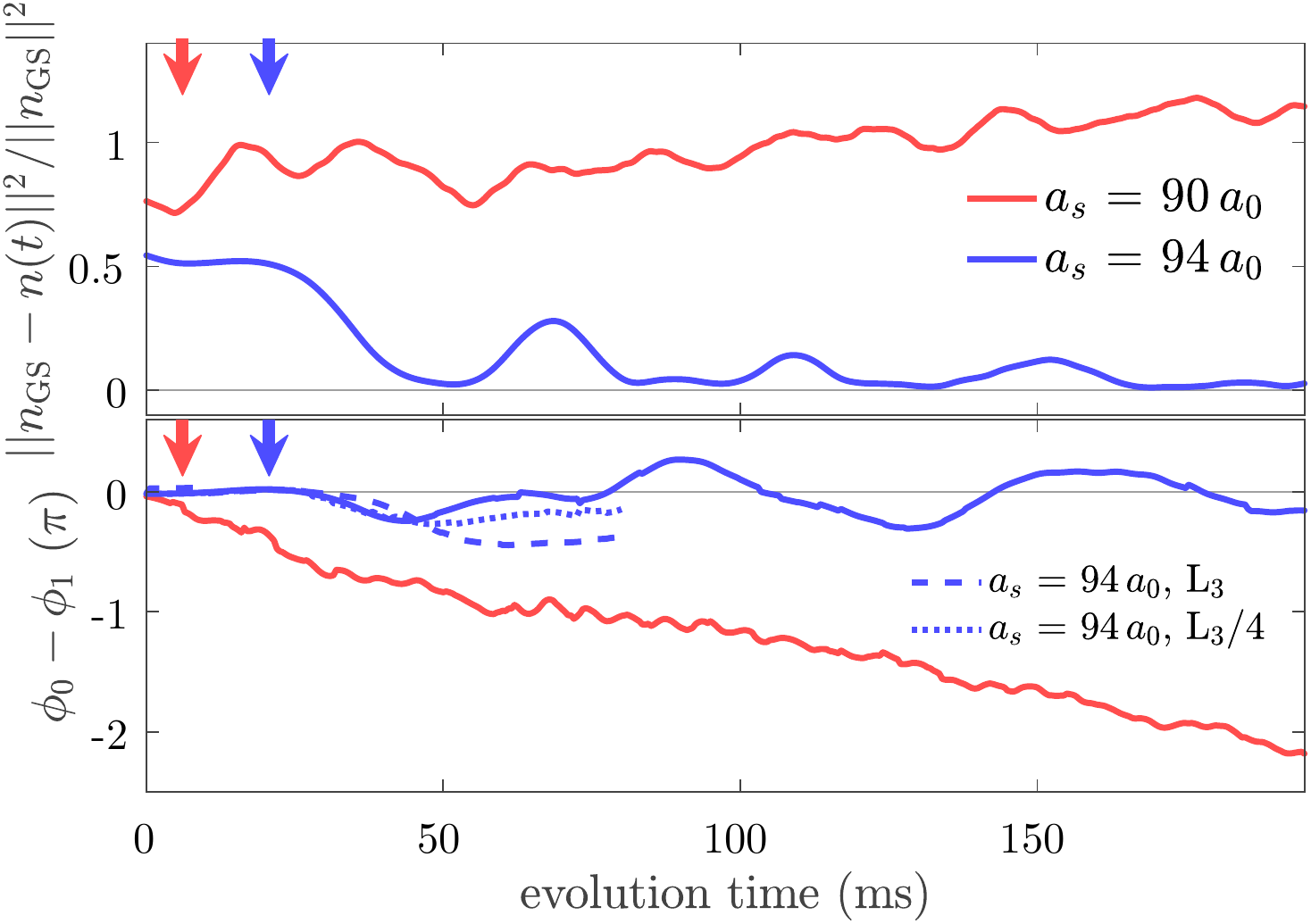}
\put(13,43){a)} \put(13,12.5){b)}
\end{overpic}
\caption{\label{fig:2} 
\textbf{Overlap of the calculated dynamical state with the ground state.} a) Time evolution of the density difference $||n_{\text{GS}}-n(t)||^2 /||n_{\text{GS}}||^2$, where $n(t)$ is the dynamical density, $n_{\text{GS}}$ is the ground state density and $||...||$ is the euclidean norm. We observe little density overlap for the isolated droplets (red) and (except for a residual breathing mode) high density overlap for the coherent droplet state (blue). b) Time evolution of the phase difference $\phi_0 - \phi_1$ between the central droplet and its nearest neighbor for the dynamical simulation. For the coherent droplet the phase difference caused during the formation process is rapidly compensated leading to a vanishing phase difference between the two neighboring droplets. On the other hand the phase difference of isolated droplets increases almost linearly after the formation, which is due to their difference in chemical potential. Including three-body losses we see a constant (experimentally measured $L_3$, dashed) or slowly decreasing ($L_3 /4$, dotted) phase difference during the lifetime of the droplets. The arrows indicate the respective formation time of the droplets for the two scattering lengths.
}
\end{figure}

To study how one can dynamically establish a phase coherent state in an experiment, we perform time dependent simulations starting from the BEC ground state at $a_{\text{s}} = 110\,a_{\text{0}}$ and then linearly ramp the scattering length to its final value within 30\,ms. A snapshot of the system after some evolution time is shown schematically in Fig.~\ref{fig:1}a, where the dynamically reached states are compared to the calculated ground states. This comparison shows that for $a_{\text{s}} = 94\,a_{\text{0}}$ the dynamically calculated wavefunction is phase-coherent and very close to the calculated ground state. In contrast to this for $a_{\text{s}} = 90\,a_{\text{0}}$, the phase of the individual droplets is different and also their number does not match the ground state prediction \cite{SupMat, Wenzel2017}. \textnew{Using shorter ramp times in the dynamical simulations, we produce states with a different number of droplets also for $a_{\text{s}} = 94\,a_{\text{0}}$. Only for ramp times $\gtrsim$\,20\,ms we produce states that are close to the ground state of the system.}

For a more \textold{detailed} \textnew{quantitative} comparison, we compare the overlap of the dynamically simulated wavefunction in real time with the ground state we get from the imaginary time evolution. We observe that incoherent droplets form soon after the end of the interaction ramp, while coherent droplets emerge slowly over several tens of milliseconds. This longer formation time for the coherent state is in agreement with a lower energy difference between this state and the initial BEC. In Fig.~\ref{fig:2}a we show the difference of the densities in the dynamical case $n(t)$ and the ground state $n_{\text{GS}}$  and Fig.~\ref{fig:2}b depicts the evolution of the phase difference between the central droplet and its nearest neighbor, which is an indicator for their phase coherence. In the density difference one can see that for the case of isolated droplets the difference is significant and actually increases with time, while for the coherent droplets the difference approaches zero after the time required for the droplets to form. The oscillations visible in the density difference after the formation correspond to a breathing mode along the droplet array. In the phase difference we observe that for the isolated droplets the phase difference increases linearly after the formation. We attribute this to the different chemical potentials of the two droplets. In contrast to this in the coherent regime we observe that the phase difference remains significantly smaller with very little variation over time. Our simulations thus reveal the existence of a state that is both density-modulated and phase coherent and can be reached dynamically by an interaction ramp into a narrow range of interaction strengths. Given the superfluid fraction and excitation spectrum that were calculated in \cite{Roccuzzo2018} we conclude that this state can be identified as a dipolar supersolid.

As a next step we include realistic three-body losses in the simulations with our experimentally measured loss coefficient $L_3 = 1.5\,\times\,10^{-40}\,\text{m}^6/\text{s}$ for $a_{\text{s}} = 94\,a_{\text{0}}$ \cite{Boettcher2019}, as well as, for reference with a lower loss rate of $L_3 / 4$. When loss is included a comparison of the densities is challenging because the ground state continuously changes with atom number. Therefore we restrict ourselves to the phase coherence shown in Fig.~\ref{fig:2}b and again observe only a small phase difference starting to form during the droplet formation process. However, this phase difference is rapidly stabilized and subsequently slowly decreases throughout the remaining lifetime of the state. An alternative way of characterizing the difference of the two wavefunctions is the fidelity, defined as $\mathcal{F} = |\left< \Psi_{\text{GS}} | \Psi(t) \right>|^2$. Using this we get a numerical value of $\mathcal{F} \approx 90\%$ after our experimental equilibration time of 15\,ms. This shows that we expect to dynamically create a state with transient supersolid properties very close to the actual ground state even in the presence of three-body losses. 

\begin{figure*}[t]
\begin{overpic}[width=0.86\textwidth]{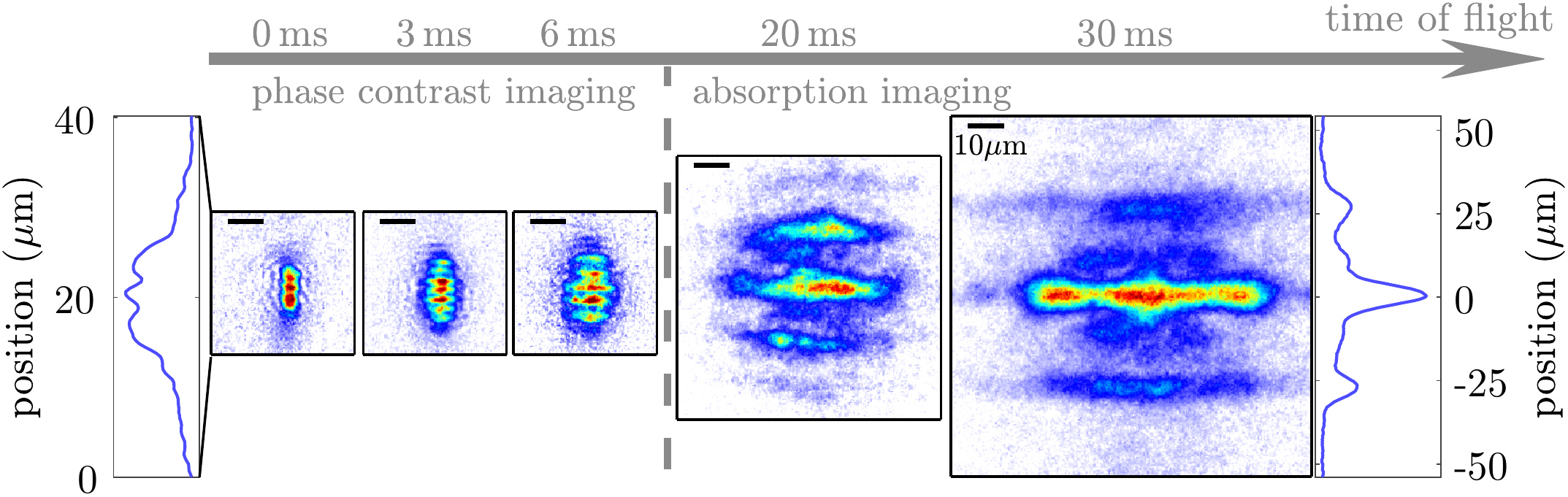}
\end{overpic}
\caption{\label{fig:3}
\textbf{Evolution from in-situ density distribution to time-of-flight interference.} On the left side we show an exemplary in-situ image together with the integrated density distribution for the phase-coherent droplet regime, revealing a clear density modulation. Towards the right side we show the expansion dynamics for different times of flight, which exhibits the characteristic increase of the fringe spacing of expanding matter waves. On the very right we show the interference pattern after 30\,ms time-of-flight together with the corresponding integrated momentum distribution. In these interference patterns a clear substructure at half the principle fringe spacing can be seen. While the principle interference peaks yield information about the nearest neighbor coherence, the additional peaks correspond to next-nearest neighbor coherence. The individual images shown result from independent experimental realizations. 
}
\end{figure*}

In order to investigate the formation of a phase-coherent droplet state experimentally, we prepare a quasi-pure dipolar BEC with approximately $4.5 \times 10^4$ $^{162}$Dy atoms at a temperature below 20\,nK in a tubular trap with trap frequencies $\omega = 2 \pi\,$(19(1), 53(1), 87(1))\,Hz and $\textbf{B}\,||\, \hat{z}$\textnew{, similar to \cite{Tanzi2018}}. We compensate the gravitational force on the atoms by ramping up a magnetic field gradient, allowing for long times of flight to probe the system. Subsequently we change the scattering length from $\sim$140\,$a_{\text{0}}$ to $\sim$110\,$a_{\text{0}}$ by ramping the magnetic field in 80\,ms closer to a double Feshbach resonances of $^{162}$Dy located around 5.1\,G \cite{Baumann2014, Boettcher2019}. In order to reach the droplet regime we subsequently ramp the magnetic field again linearly within 30\,ms to the final scattering length in the range between 89\,$a_{\text{0}}$ and 98\,$a_{\text{0}}$ \footnote{Due to the uncertainty in the measurements of the background scattering length all experimentally quoted scattering lengths used throughout this work exhibit an uncertainty on the order of 15\%. For more information on the determination of the scattering length see the Supplemental Material}. We then hold the atoms for 15\,ms at this field in order to equilibrate. Finally, we probe the resulting state either in-situ using far detuned phase-contrast imaging, or after time-of-flight using absorption imaging. As we are in or close to a regime where droplets are self-bound we boost the time-of-flight expansion velocity by ramping up the scattering length to \textold{$\sim\,a_{\text{bg}}$} \textnew{$\sim$140\,$a_0$} within 100\,$\upmu$s just before the release of the atoms from the trapping potential. This has the additional advantage that density rapidly decreases, minimizing interaction effects in the expansion. Moreover, it acts as a zoom greatly increasing the extend of the interference patterns, and thus, giving access to subtle features beyond simple nearest-neighbor phase coherence. As exemplified in Fig.~\ref{fig:3}, our experiment thereby reveals both the in-situ density modulation as well as the interference pattern of multiple matter waves emerging after time of flight. 

\begin{figure}[t]
\begin{overpic}[width=0.46\textwidth]{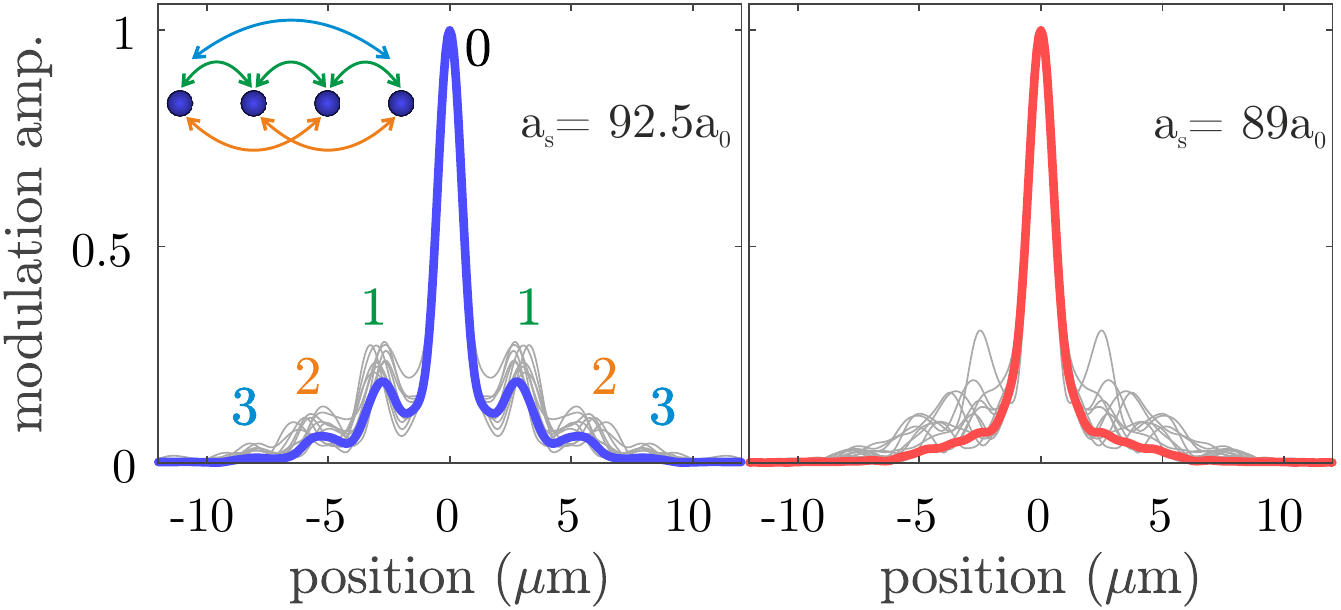}
\put(50,40){a)} \put(94,40){b)}
\end{overpic}
\caption{\label{fig:4} 
\textbf{Evaluation of the coherence between neighboring droplets.} Absolute value of the Fourier transform of the integrated interference patterns after 30\,ms time of flight, showing signs of the multiple frequencies of the interference due to  nearest-neighbor, next-nearest neighbor and even higher-order coherence. This is schematically shown in the inset of Fig.~\ref{fig:4}a. The gray lines correspond to single shot realizations and the blue lines to the mean of all available realizations for a given atom number for $a_{\text{s}} = 92.5\,a_{\text{0}}$ (a, blue) and $a_{\text{s}} = 89\,a_{\text{0}}$ (b, red). In the latter case the position of the side peaks changes randomly (b), while in the coherent droplet regime (a) the side peaks appear at the same position in every realization and show very little variance in their amplitude. This provides clear evidence for the phase coherence between the droplets (a), with the variance of the peak height of the first side peak corresponding to nearest neighbor coherence, the second peak  corresponding to next-nearest neighbor coherence and the third to next-next-nearest neighbor coherence. The shown exemplary Fourier transforms correspond to the red points in Fig.~\ref{fig:5}b and c.
}
\end{figure}

\begin{figure*}[t]
\begin{overpic}[width=0.98\textwidth]{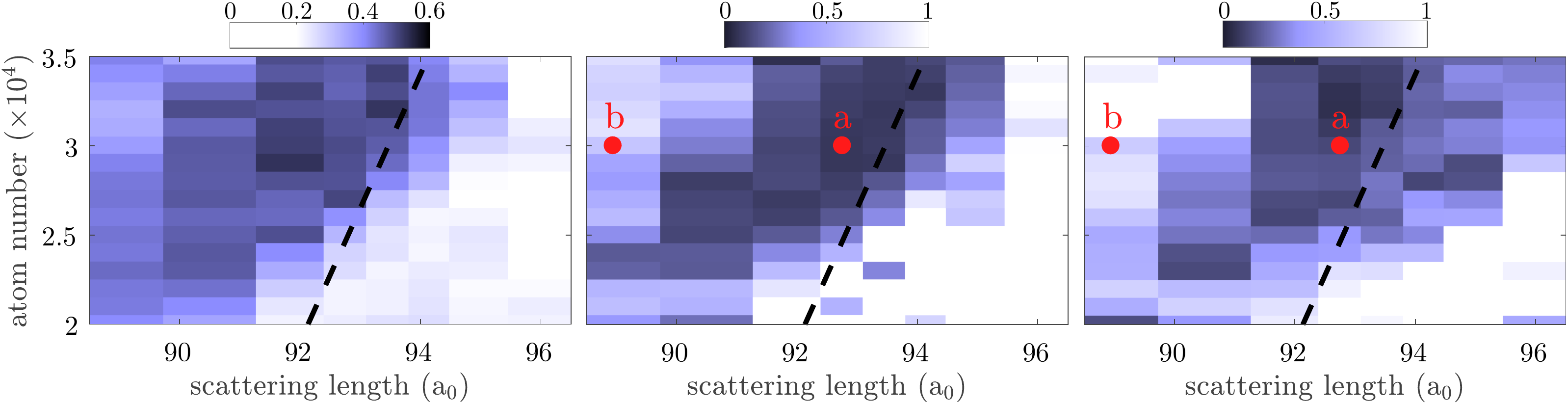}
\put(0,24){a)} \put(37,24){b)} \put(69,24){c)}
\end{overpic}
\caption{\label{fig:5} 
\textbf{In-situ modulation and phase coherence reveal signatures of the theoretical phase diagram.} Spectral weight of the observed in-situ modulation (a), nearest neighbor coherence (b) and next-nearest neighbor coherence (c). Only in the range where we observe a density modulation in (a) we also see interference patterns emerging in time of flight (b, c). For a narrow range of the contact interaction strength we see clear evidence for phase coherence up to the next-nearest neighbor. The red points labeled a and b correspond to the exemplary Fourier transforms shown in Fig.~\ref{fig:4}a and b. The dashed black lines are the same guide to the eye that was shown in Fig.~\ref{fig:1}c.
}
\end{figure*}

First we analyze in more detail how the in-situ density distribution changes for different atom numbers and final scattering lengths. Above a certain atom number threshold we observe the appearance of multiple droplets aligning along the weak axis of the trap. \textold{However due}\textnew{Experimentally this density modulation takes about 5-10\,ms to develop. We attribute this faster formation time compared to the numerical simulations to fluctuations due to residual excitations and finite temperature, which can seed the underlying instabilities driving the phase transition. Due} to our finite imaging resolution of 1\,$\upmu$m, the smaller size of the droplets and the imaging aberrations arising thereupon, we cannot reliably extract the number of droplets or the overlap between droplets and the background BEC directly from our images. This resolution limited imaging also leads to a larger uncertainty in the extracted atom number compared to the data obtained in time-of-flight. As in our previous work \cite{Kadau2016}, we therefore use the absolute value of the Fourier transform of the integrated in-situ density to identify images with a density modulation. To this end, we compare the spectral weight of \textold{finit} \textnew{finite} momentum contributions to the weight of the central peak in momentum space \cite{SupMat}. This ratio is plotted as a function of scattering length and atom number in Fig.~\ref{fig:5}a. Typically, every coordinate is an average of a few experimental runs with 80 realizations in total for every scattering length and the different atom numbers have been realized by binning our experimental data. We observe that a density modulation appears above a certain atom number threshold and compare this to the boundary from the theoretical phase diagram. With this we can clearly map out a region showing a density-modulated state in the experiment. For small scattering lengths and high atom number we see the modulation amplitude decreasing again, which is caused by a washing out of the mean distribution due to fluctuations in the number of droplets and therefore also their spatial separation.

Next we study the phase coherence of the realized droplet states via interference after 30\,ms time-of-flight. For our parameters we typically realize an array of several droplets. In contrast to the well known interference of two BECs \cite{Davis1995} our situation thus leads to a more complex interference pattern with multiple frequencies \cite{Hadzibabic2004, SupMat}. For the evaluation of our data we therefore again turn to the absolute value of the \textold{Fourier transformation} \textnew{Fourier transform} of the integrated image after time-of-flight. Note, that this Fourier transform of the time-of-flight density, rather than the wavefunction, does not yield again the initial in-situ wavefunction but rather provides information about the individual frequency components of the interference pattern. Examples of this for scattering lengths of $a_{\text{s}} = 92.5\,a_{\text{0}}$ and $a_{\text{s}} = 89\,a_{\text{0}}$ are shown in Fig.~\ref{fig:4}a and b, respectively. The gray lines are single-shot realizations, while the blue and red lines show the mean of all available realizations for a given atom number and scattering length, with more than 300 realizations for every scattering length in total. 

We observe a clear difference between the two scattering lengths, caused by two distinct effects. The data for $a_{\text{s}} = 92.5\,a_{\text{0}}$ can be identified with coherent droplets and show stable side peaks at the same position in every realization. The data for $a_{\text{s}} = 89\,a_{\text{0}}$ correspond to incoherent droplets and shows strong fluctuations from realization to realization. The change in the side peak positions for the incoherent droplets is caused by a varying number of droplets and therefore a different initial separation from shot to shot, while the stable position of the side peaks for the coherent droplets means that the initial state is very reproducible. In particular, as in the usual double-slit interference, the height of the Fourier peaks are a measure of the interference contrast and their shot-to-shot variation thus encodes the phase coherence of the system. In our case, the individual Fourier peaks characterize nearest-neighbor, next-nearest neighbor and even higher-order coherences \cite{SupMat}. If the initial droplets are phase-coherent the side peaks are expected to always exhibit the same height and therefore the corresponding variance should be low. On the other hand, if the initial droplets are not phase-coherent the peak height should fluctuate from realization to realization and therefore we should observe an increase in the variance of the peak height. Taken together these two effects -- fluctuating droplet number and, hence, fringe spacing, as well as incoherent phases between independent droplets -- wash out the mean distribution for $a_{\text{s}} = 89\,a_{\text{0}}$, while the mean distribution for $a_{\text{s}} = 92.5\,a_{\text{0}}$ shows clear side peaks. This already provides clear evidence for the phase coherence between the droplets in the latter case, with the first side peak (labeled as 1 in Fig.~\ref{fig:4}a) corresponding to nearest neighbor coherence, the second peak (labeled as 2) corresponding to next-nearest neighbor coherence. For atom numbers high enough to yield more than three droplets we can even observe a small signal of next-next-nearest neighbor interference (labeled as 3). 

To quantify the phase coherence we therefore calculate the variance of the height of the Fourier transform side peaks (1 and 2 in Fig.~\ref{fig:4}a) normalized to the peak height \cite{SupMat}. We show these ratios in Fig.~\ref{fig:5}b and c for different atom numbers and final scattering lengths. The results are in very good agreement with our in-situ results in Fig.~\ref{fig:5}a, as well as the theoretical phase diagram shown in Fig.~\ref{fig:1}c. Again we find a sharp phase boundary where the multiple droplet becomes energetically favorable compared to the BEC state. Below this boundary in the respective atom number, we observe no interference and therefore no signal in the Fourier transform. Above this threshold we always see interference, however only for a small range of contact interactions the coherence between droplets is present. The observed boundaries in all these plots are in agreement with the simulated ground state phase diagram in Fig.~\ref{fig:1}c. Combining the in-situ with the interference results reveals the signatures of the theoretical phase diagram and we see that for a small range of the contact interaction strength there exists a phase of the system showing both a density modulation as well as phase coherence and therefore the hallmark properties of a supersolid state of matter.

In conclusion we have shown theoretically and experimentally that for a narrow range of interaction strengths our dipolar quantum gas of $^{162}$Dy atoms exhibits a state that is both density-modulated and phase-coherent. \textold{This} \textnew{Together with the dynamical study of the phase coherence in \cite{Tanzi2018}, this} observation is the first step towards the realization and identification of a dipolar supersolid, where in contrast to previous works \cite{Leonard2017, Leonard2017a, Li2017} the self-organized density modulation is induced by the intrinsic interactions. In order to finally \textold{proof} \textnew{prove} the supersolid character of the observed state beyond the phase coherence demonstrated in this work, an experimental proof of phase rigidity, and hence genuine superfluidity, is required. As a next step we therefore plan to investigate the two types of collective excitations, the phonon and phase modes. Another important aspect we plan to study is to extend the lifetime of the observed states which is currently limited due to three-body losses to approximately 20\,ms. This could be accomplished by identifying a region with lower losses in the rich Feshbach spectrum of $^{162}$Dy \cite{Baumann2014, Frisch2014, Maier2015, Maier2015a}.

\begin{acknowledgments}
We acknowledge insightful discussions with H.P. B\"{u}chler, I. Ferrier-Barbut, F. Ferlaino, L. Santos and A. Pelster. This work is supported by the German Research Foundation (DFG) within FOR2247 under Pf381/16-1, Pf381/20-1, and FUGG INST41/1056-1. T.L. acknowledges support from the EU within Horizon2020 Marie Sk\l odowska Curie IF (Grants No.~746525 coolDips), as well as support from the Alexander von Humboldt Foundation through a Feodor Lynen Fellowship.
\end{acknowledgments}

\appendix

\section{Experiment}

The complete experimental sequence is described in more detail e.g. in \cite{Kadau2016}. In short we create Bose-Einstein condensates of $^{162}$Dy in a crossed optical dipole trap (cODT) made up of two laser beams with $\lambda_{\text{cODT}} = 1064\,$nm. After the creation of the BEC we change the trap to $\omega = 2 \pi\,$(19(1), 53(1), 87(1))\,Hz within 20\,ms.

Imaging is performed along the magnetic field axis ($\hat z$ axis) using a microscope objective that allows us to reach a resolution of about $1\,\upmu$m. We can image the atoms either with phase-contrast imaging in-situ or with resonant absorption imaging in time of flight. Both techniques result in similar atom numbers with an overall uncertainty of $<10\%$. To keep the atomic cloud in the focus of our objective we apply a magnetic field gradient to compensate gravitational forces. This applied gradient leads to a shift of the magnetic field by $-428\,$mG at the position of the atoms, which we compensate by ramping up the amplitude of the magnetic offset field at the same time as the gradient.

\textnew{We observe the phase-coherent droplets (point a in Fig.~\ref{fig:5}b of the main text) at a magnetic field of 5.266(1)\,mG, calibrated using RF loss spectroscopy. The whole scattering length range shown in Fig.~\ref{fig:5} corresponds to a magnetic field amplitude range of 5.261(1) to 5.271(1)\,mG.}
        
\section{Feshbach resonances}  

For the measurements in this work we use a specific double Feshbach resonance of $^{162}$Dy from the rich spectrum of resonances \cite{Baumann2014, Frisch2014, Maier2015, Maier2015a} in order to control the short-range contact interaction. The resonances are located at fields of $B_{\text{1}} = 5.126(1)\,\text{G}$ and $B_{\text{2}} = 5.209(1)\,\text{G}$ with widths of $\Delta B_{\text{1}} = 35(1)\,\text{mG}$ and $\Delta B_{\text{2}} = 12(1)\,\text{mG}$, respectively. Details about the measurement and calibration of these resonances will be presented in future work \cite{Boettcher2019}. Additionally we also include a broader resonance at $B_{\text{3}} = 21.95(5)\,\text{G}$ with a width of $\Delta B_{\text{3}} = 2.4(8)\,\text{G}$ \cite{Lucioni2018} into our considerations, since this resonance still has a small effect on the scattering length in the magnetic field range considered in this work. This allows us to calculate the scattering length $a_{\text{s}}$ from the magnetic field we measure, with only the background scattering length as a free parameter. $^{162}$Dy features a rather high background scattering length $a_{\text{bg}} = 140(20)\,a_{\text{0}}$ \cite{Tang2018, Tang2015a, Tang2016}, which means that away from the resonances the sample is contact-dominated, while closer to the zero-crossing of the resonance we can create a dipolar-dominated sample. In the range of scattering length studied in this work our typical field stability leads to an uncertainty of $\sim 1\,a_0$. However, due to the uncertainty in the measurements of the background scattering length, all scattering lengths used throughout this work exhibit an uncertainty on the order of 15\%.

Another important parameter controlled by the used Feshbach resonances is the three-body loss coefficient $L_3$. We measure a three-body loss coefficient of $L_3 = 1.33 \times 10^{-41}\,\text{m}^6/\text{s}$ away from the Feshbach resonances for a BEC of $^{162}$Dy. Closer to the resonances the loss coefficient increases rapidly \cite{Boettcher2019} leading to increased losses and therefore shorter lifetimes of the observed coherent droplets. For the scattering length where we observe the coherent droplets the losses are already enhanced by about a factor of 9 compared to the value away from the resonance. By using a different Feshbach resonance with lower loss rates from the complicated spectrum of resonances one could therefore greatly increase the lifetime of the observed states.

\section{In-situ evaluation}

\begin{figure}[t]
\begin{overpic}[width=0.48\textwidth]{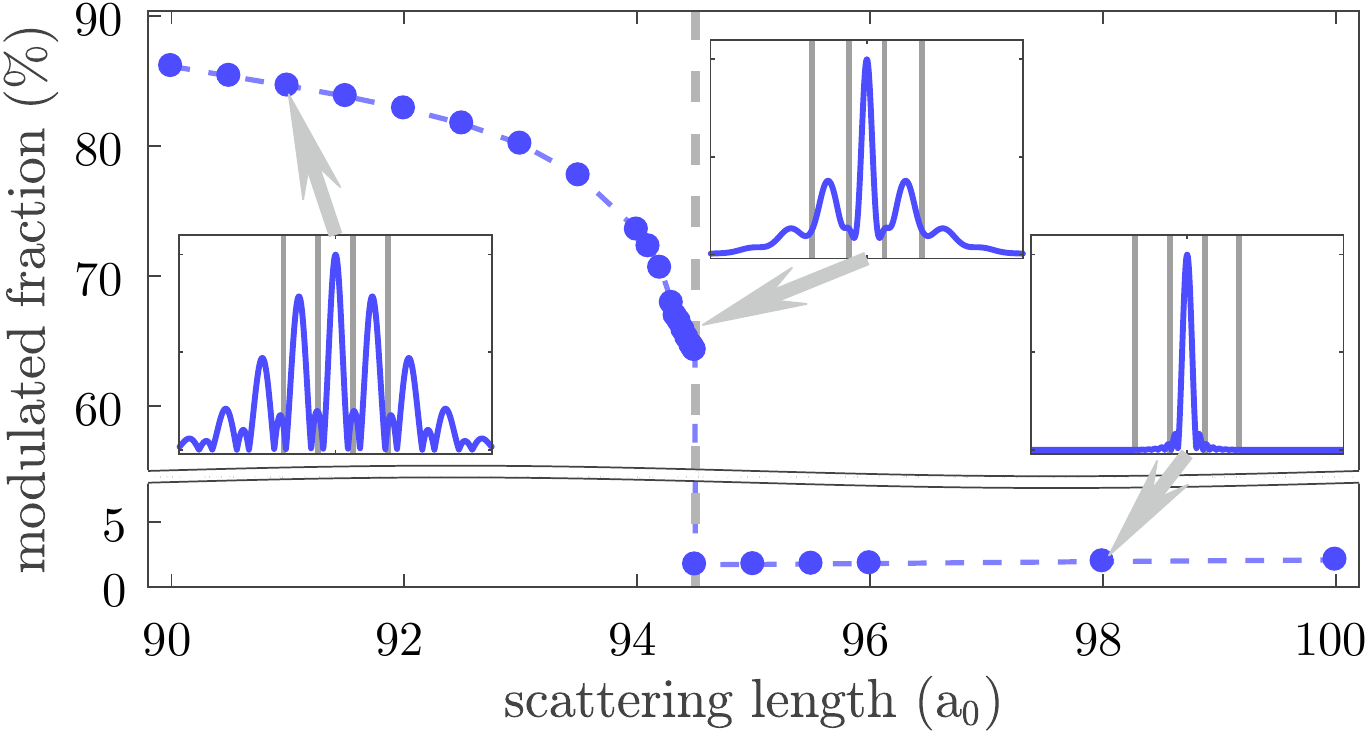}
\end{overpic}
\caption{\label{fig:SupMat1} 
\textbf{Spectral weight across the phase transition from BEC to immersed droplets to isolated droplets.} Alternatively to Fig.\ref{fig:1}b of the main text, we can also quantify the transition by calculating the spectral weight $SW$ of the first momentum peak compared to the zeroth order according to Eq.~\ref{Eq:SpectralWeight}. This is done for the calculated ground states for a scattering length of $a_{\text{s}} = 94\,a_0$ and containing $3.5 \times 10^4$ atoms. The insets show the absolute value of the Fourier transform exemplary for points in the three different regimes.
}
\end{figure}

As explained in the main text we observe the appearance of an array of multiple droplets above a certain atom number threshold. The small theoretical size of the droplets compared to our imaging resolution leads to imaging aberrations. These aberrations mean that directly fitting the data, as well as directly counting the droplets is not reliable. We therefore quantify the density modulation by analyzing its contribution to the absolute value of the Fourier transform of the integrated images.

To show the general principle of this analysis we first demonstrate it on the simulated ground states that we used in Fig.~\ref{fig:1}b of the main text to show the phase transition from BEC to coherent droplets to isolated droplets. For a BEC the absolute value of the Fourier transform is peaked around zero with almost no contribution at higher momentum. Crossing the boundary into the droplet phase we observe a pronounced local maximum at a finite momentum. With decreasing scattering length we observe higher order peaks appearing that get more and more pronounced. To quantify the transition we determine the spectral weight $SW$ at finite momentum instead of the ratio of the overlap used in the main text. For the simulated ground states we therefore calculate the weight of the first order momentum peak and compare it to the central peak, by summing up the corresponding signal $S(k)$ in the range indicated by the gray lines in Fig.~\ref{fig:SupMat1}.

\begin{equation} \label{Eq:SpectralWeight}
SW = \sum_{|k|= 0.7 \upmu\text{m}^{-1}}^{2.1 \upmu\text{m}^{-1}}S(k) \quad /\quad \sum_{|k|< 0.7  \upmu\text{m}^{-1}}S(k)
\end{equation}

This spectral weight is shown in Fig.~\ref{fig:SupMat1} across the phase transition from a BEC to quantum droplets. We observe a distinct jump in the spectral weight at the phase transition. Interpreting this as an order parameter this discontinuity, similar to the measure of the overlap used in the main text, is an indication of a first-order phase transition.

\begin{figure}[t]
\begin{overpic}[width=0.48\textwidth]{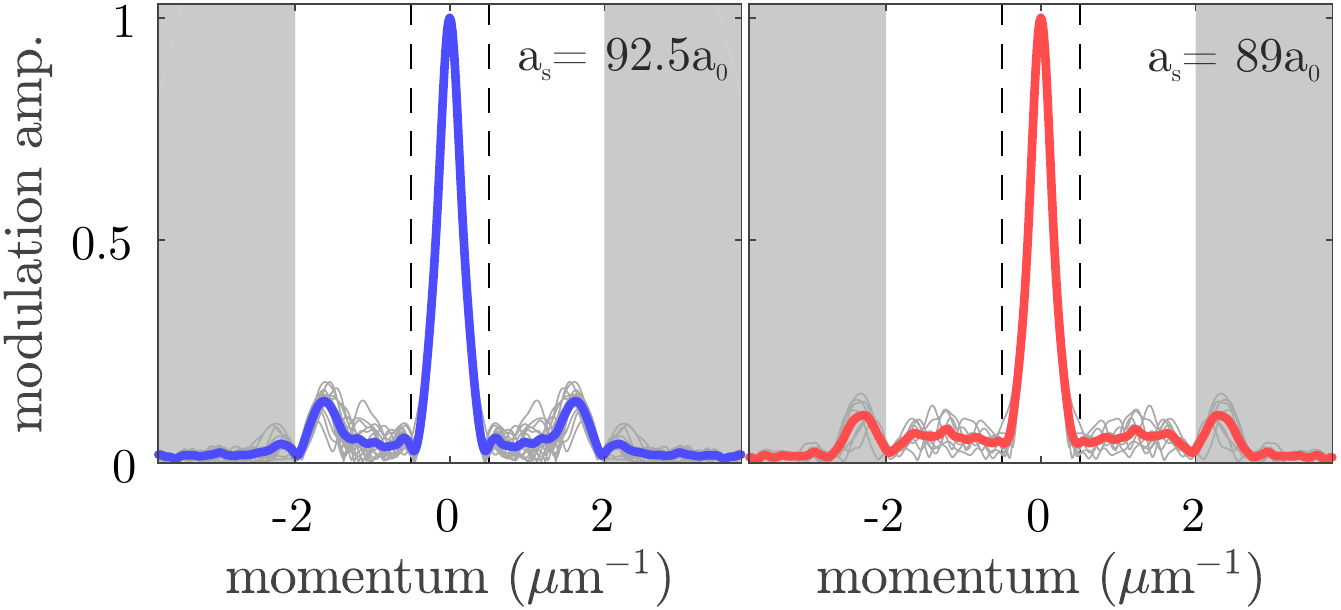}
\put(15,40){a)} \put(57,40){b)} 
\end{overpic}
\caption{\label{fig:SupMat2} 
\textbf{Evaluation of the density modulation in the in-situ images.} Absolute value of the Fourier transform of the integrated in-situ images, showing clear indications of a density modulation with finite momentum. The gray lines correspond to single shot realizations and the blue and the red lines correspond to the mean of all available realizations for a given atom number for $a_{\text{s}} = 92.5\,a_{\text{0}}$ (a) and $a_{\text{s}} = 89\,a_{\text{0}}$ (b), respectively. The dashed gray lines indicate the boundary between main peak and finite momentum distribution used for the calculation of the spectral weight. The gray area is excluded from the analysis due to our imaging aberrations.
}
\end{figure}

For Fig.~\ref{fig:5}a of the main text we used this analysis in order to map out the phase boundary between BEC and the density modulated arrays of quantum droplets. In Fig.~\ref{fig:SupMat2}a and b we show the absolute value of the Fourier transform of the integrated in-situ images exemplary for the case of coherent droplets ($a_{\text{s}} = 92.5\,a_{\text{0}}$) and incoherent droplets ($a_{\text{s}} = 89\,a_{\text{0}}$), respectively. Differently than for the simulated ground states we do not observe any higher order peaks in our experimental data, due to the finite imaging resolution. In order to calculate the spectral weight for the experimental data we sum up the signal with $|k| < 0.5 \upmu\text{m}^{-1}$ for the main peak and with $0.5 \upmu\text{m}^{-1} < |k| < 2 \upmu\text{m}^{-1}$ for the finite momentum contribution. The boundary between the two ranges is indicated by the dashed gray lines in Fig.~\ref{fig:SupMat2}a and b. For large $|k|$ we also put a boundary due to the imaging aberrations we observe. These imaging aberration are more pronounced for the isolated droplets compared to the coherent droplets.

\section{Interference of an array of phase independent matter waves}

The interference of two waves, whether they are light waves after a double slit or matter waves from two separate sources \cite{Andrews1997, Hadzibabic2004}, is approximately described by the well-known interference pattern

\begin{equation}\label{Eq:DoubleSlit}
I(z) \propto e^{-\frac{2 z^2}{Z_{0}^2}} \; \left( 1 + A_1 \; \cos \left( \Phi_1 + \frac{2\pi z}{D} \right) \right) \,.
\end{equation}
In this $A_1$ is the contrast of the interference pattern and $\Phi_1$ the relative phase of the two waves. In the case of matter waves initially confined in a harmonic trap with trapping frequency $\omega_z$, the interference fringe spacing given by $D = ht/(md)$, with $d$ the separation of the initial traps, $t$ the time-of-flight, $Z_0 = \hbar t/(ml)$ the size of the matter waves in time-of-flight, $m$ the atomic mass and $l = \sqrt{ \hbar/(2m\omega_z)}$ the width of the initial density distribution. Here, we have assumed long times of flight $t$. 

In comparison to this, a more complex interference pattern characterized by multiple frequencies arises from the interference of light waves passing through a multi-slit arrangement or for multiple interfering matter waves. Here, our aim is to characterize the phase coherence between these individual matter waves. Following \cite{Hadzibabic2004} we can write the interference pattern $I(z)$ for an array of $N$ independent matter waves according to

\begin{equation} \label{Eq:MultiSlit}
I(z) \propto e^{-\frac{2 z^2}{Z_{0}^2}} \; \left( 1 + \sum_{n=1}^{N-1} \; A_n \; \cos \left( \Phi_n + \frac{2\pi n z}{D} \right) \right) \;,
\end{equation}
where the amplitudes $A_n$ and the phases $\Phi_n$ are given by the modulus and the argument of a sum of complex numbers
\begin{equation} \label{Eq:AmpPhase}
\frac{2}{N} \sum_{j=n+1}^{N} \; e^{i(\phi_j - \phi_{j-n})} \;.
\end{equation}

In this $\phi_j$ is the phase of the single matter wave at position $j$. The contrast amplitudes $A_n$ and the phases $\Phi_n$ of the interference pattern are thus defined by the phase differences of all matter wave pairs initially separated by the distance $nd$.  Here the index $n$ runs over all the possible pairs of interfering waves, starting from the nearest neighbors with $n = 2$ and ending with the interference of the two outermost waves with $n = N-1$. This is schematically indicated in the inset of Fig.~\ref{fig:4}a of the main text. 

The result in Eq.~\ref{Eq:MultiSlit} can be obtained from the general expansion of matter waves, assuming that they all contain the same average number of particles and that the atoms do not interact during the expansion. 

For only two interfering matter waves the interference pattern simplifies to Eq.~\ref{Eq:DoubleSlit}, while for infinitely many interfering waves it corresponds to the interference pattern of a grating given by
\begin{equation}
I(z) \propto \frac{1}{N} \; \frac{\sin^2(N \pi z /D)}{\sin^2(\pi z /D)} \, .
\end{equation}

\begin{figure}[t]
\begin{overpic}[width=0.48\textwidth]{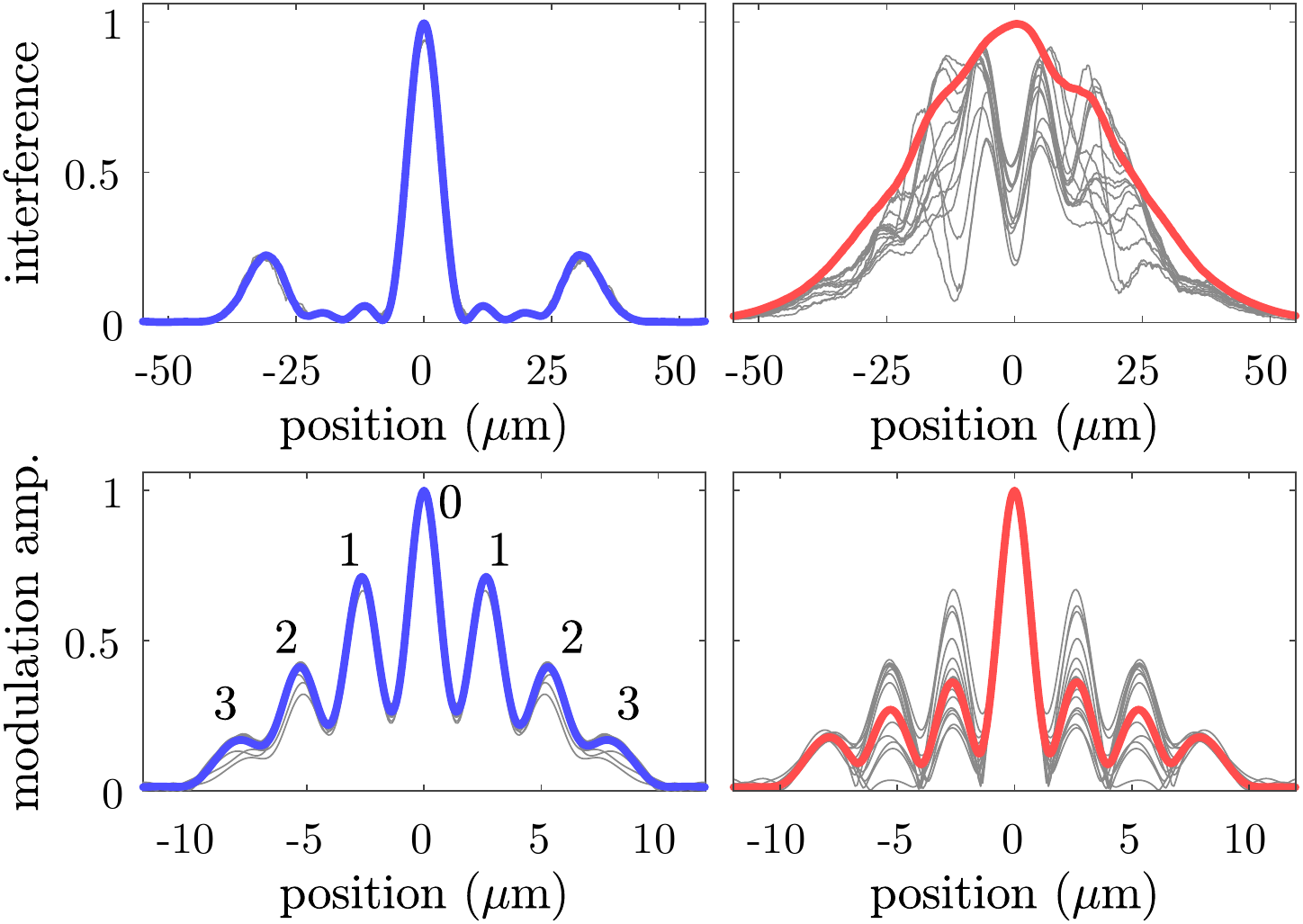}
\put(13,66){a)} \put(58,66){b)} \put(13,30){c)} \put(58,30){d)}
\end{overpic}
\caption{\label{fig:SupMat3} 
\textbf{Simulated interference patterns and corresponding Fourier transforms for coherent and incoherent matter waves.} We use Eq.~\ref{Eq:MultiSlit} to simulate the interference patterns of four coherent (a) and incoherent (b) matter waves using the experimentally observed fringe spacing and cloud width. Here, the blue and red lines denote mean interference patterns and mean Fourier transforms for the coherent and incoherent case, respectively. Gray lines represent exemplary single shot realizations. While the interference pattern are perfectly reproducible from shot-to-shot for the coherent waves, the incoherent ones exhibit strong fluctuations. Using the Fourier transform we see stable side peaks with always the same height for the coherent interference (c), while the large shot-to-shot variance of the peak heights in the incoherent case (d) leading to a lower mean peak height. Note that experimentally, the assumption of similar atom numbers that underlies Eq.~\ref{Eq:MultiSlit} is clearly not valid. This leads to a decrease of amplitude for the side peaks, which can also be observed in our eGPE simulations. 
}
\end{figure}

\begin{figure*}[t]
\begin{overpic}[width=0.98\textwidth]{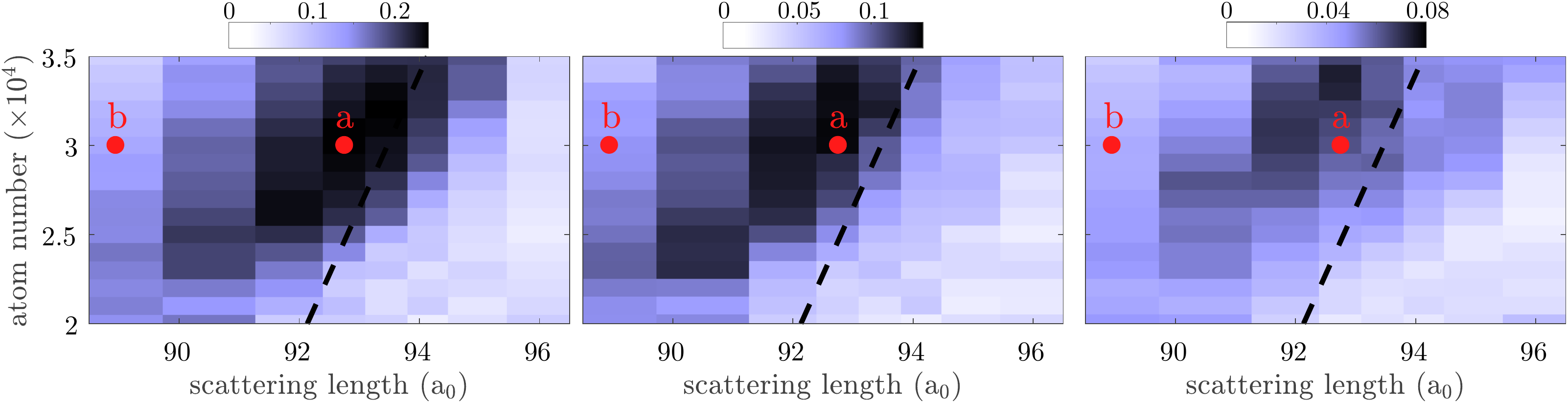}
\put(0,24){a)} \put(37,24){b)} \put(69,24){c)}
\end{overpic}
\caption{\label{fig:SupMat4} 
\textbf{Phase coherence using the weight of the side peaks.} Weight of the side peaks compared to the main peak of the Fourier transforms. For a narrow range of the contact interaction strength we see a larger signal, which we interpret as phase coherence up to the next-next-nearest neighbor. The phase diagrams are in excellent agreement with the ones presented in Fig.~\ref{fig:5}b,c of the main text using the variance of the peak height as indicator. The red points labeled a and b correspond to the exemplary Fourier transforms shown in Fig.~\ref{fig:4}a and b. The dashed black lines are a guide to the eye and are taken from Fig.~\ref{fig:1}c of the main text.
}
\end{figure*}

Using Eq.~\ref{Eq:MultiSlit} we can generate 300 idealized interference patterns for four phase coherent matter waves (Fig.~\ref{fig:SupMat3}a), or for four matter waves with randomly distributed phases (Fig.~\ref{fig:SupMat3}b). The blue solid line corresponds to the mean interference pattern in the coherent case ($\Phi_n = 0$) and the red solid line to the mean interference pattern in the incoherent case (random distributed $\Phi_n$). Examples of single shot realizations are shown in gray. We can see that the interference pattern is perfectly reproducible in the case of coherent droplets, while there are strong fluctuations in the incoherent case, which lead to a washed out mean distribution even if the single-shots show a large contrast. This means that the appearance of high contrast interference pattern alone is not enough to proof phase coherence.

To analyze this in more detail we can use the same evaluation with the Fourier transform as for the experimental data. For Eq.~\ref{Eq:MultiSlit} this leads to the spectra shown in Fig.~\ref{fig:SupMat3}c and d, where the peaks of the absolute value of the Fourier transform are located at multiples of $d$. The height of these side peaks are proportional to $A_n$. Again the blue and red line correspond to the respective mean distribution, while the gray lines are exemplary single realizations. As expected from Eq.~\ref{Eq:AmpPhase} the single realizations for the phase coherent waves show stable side peaks with very low variance in the peak heights for different realizations, while the peak heights for the incoherent matter waves show strong fluctuations. The Fourier transform - and specifically the variance of the peak heights - thus allows us to individually quantify the coherences of order $n$.

\section{Revealing the experimental signatures of the phase diagram}

In Fig.~\ref{fig:5}b and c of the main text we show the variance of the height of the side peaks as a measure of the phase coherence between neighboring droplets, as it was explained in the previous section and in Fig. \ref{fig:SupMat4}. In order to get a result independent of the peak height, we normalize it to the corresponding mean peak height
\begin{equation}
\frac{1}{(N-1) \left< p \right>^2} \; \sum_{i=1}^N \; \left| p_i  - \left< p \right> \right|^2
\end{equation} 
with $N$ the number of realizations, $p_i$ the peak height of the $i$-th realization and $\left< p \right>$ the mean peak height. In this evaluation we subtract the background signal which we observe in the Fourier transforms even in the absence of interference from the mean peak height.

Instead of looking at the variance of the heights of the side peaks in the Fourier transforms (see Fig.~\ref{fig:5}b and c of the main text), we can also directly study the peak height as an indicator for the phase coherence. To make the analysis more robust, rather than directly identifying a peak height in the noisy experimental data, we calculate the weight of the mean distribution around the side peaks and compare this to the weight of the main peak. We show this ratio in Fig.~\ref{fig:SupMat4}a-c for different atom numbers and final scattering lengths, up to the signal resulting from the next-next-nearest neighbor interference. The black dashed line is again the guide to the eye from Fig.~\ref{fig:1}c of the main text. These results are in very good agreement with the results obtained through the analysis of the variance in the peak heights that are shown in Fig.~\ref{fig:5}b-c of the main text, as well as the theoretical phase diagram shown in Fig.~\ref{fig:1}c of the main text.

\textnew{\section{Hysteresis in the numerical simulations}}

\begin{figure}[t]
\begin{overpic}[width=0.48\textwidth]{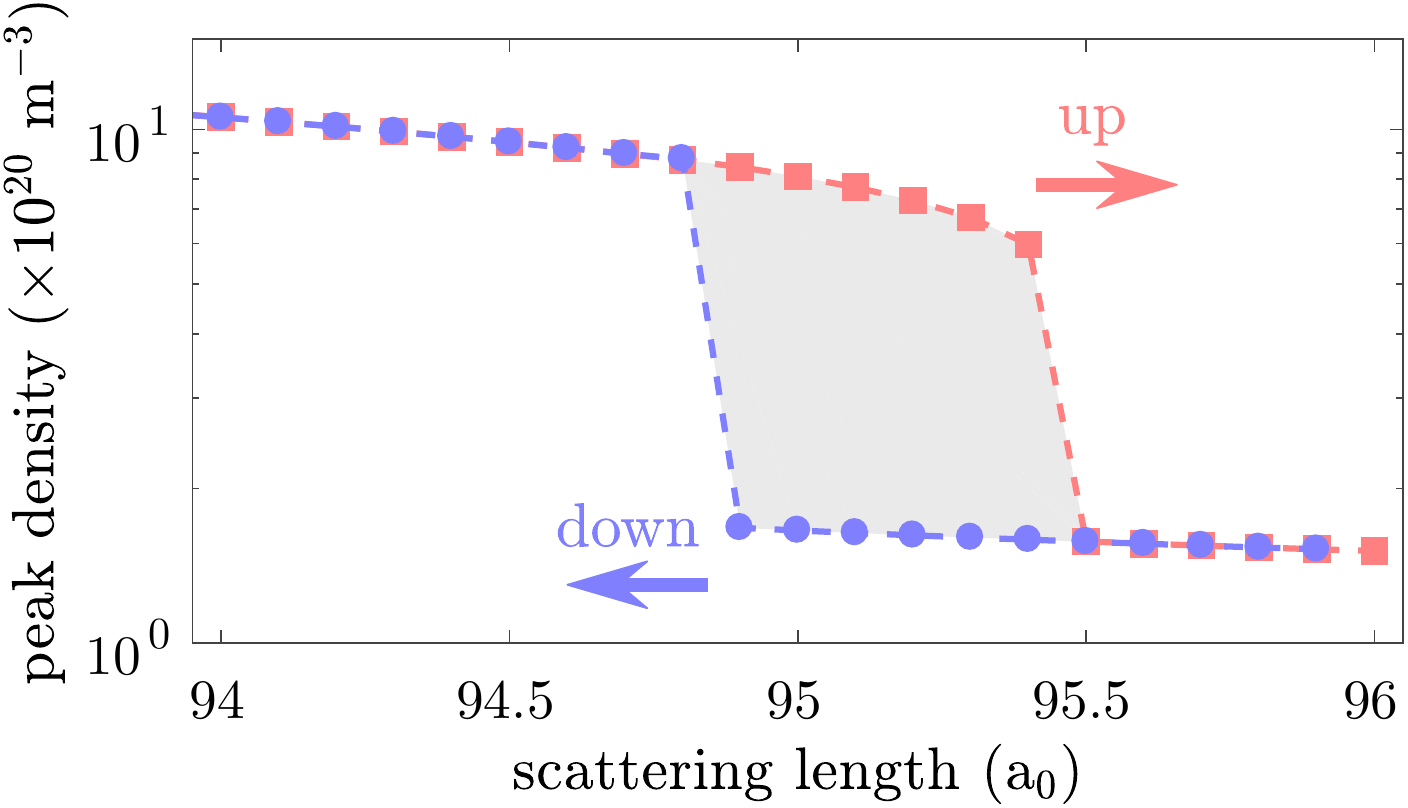}
\end{overpic}
\caption{\label{fig:SupMat5} 
\textnew{\textbf{Hysteresis of the phase transition in the numerical simulations.} We numerically solve the eGPE using imaginary time propagation to calculate the ground states, once with a regular BEC as an initial wavefunction (blue, labelled "down") and once with a density modulated, multiple droplet initial wavefunction (red, labelled "up"). We observe a hysteretic behavior depending on the direction from which we approach the transition. These simulations are carried out for $3.5 \times 10^4$ $^{162}$Dy atoms confined in the experimentally used trap.
}}
\end{figure}

\textnew{The observed discontinuity in the measure of the overlap of neighboring droplets in Fig.~\ref{fig:1}b of the main text is an indication of a first-order phase transition. This first-order nature of the transition is further evidenced by the observation of hysteresis in \cite{Tanzi2018, Kadau2016}.}

\textnew{To further study the nature of the transition, we numerically solve the eGPE using imaginary time propagation \cite{Wenzel2017} to calculate the ground states, once with a regular BEC as an initial wavefunction (labelled "down" in Fig.~\ref{fig:SupMat5}) and once with a density modulated, multiple droplet initial wavefunction (labelled "up"). These simulations are performed for $3.5 \times 10^4$ $^{162}$Dy atoms in a harmonic potential with trapping frequencies $\omega~=~2\pi\,$(18.5,~53,~81)\,Hz similar to \cite{Tanzi2018}. Again we recover the three distinct regimes of a regular BEC and the coherent and incoherent quantum droplets depending on the contact interaction strength. However, we observe that the boundary between BEC and density modulated droplets is shifted depending on the direction from which we approach the transition. This hysteretic behavior is shown in Fig.~\ref{fig:SupMat5}, where we have used the peak density as a simple probe to identify the transition point. Note that while the peak density can be used to identify the transition between BEC and quantum droplet regime, it can not distinguish between incoherent and coherent droplets. This is because the peak density is significantly higher for the quantum droplets compared to the BEC, but it is similar for the two different droplet regimes.
}

\bibliography{paper}

\end{document}